\documentclass[twocolumn,preprintnumbers,amssymb,amsmath,superscriptaddress,aps,prd,nofootinbib]{revtex4}[10pt]

\pdfoutput=1

\usepackage{color}
\usepackage{graphicx}

\begin{document} 

\newcommand{\beq}{\begin{eqnarray}}
\newcommand{\eeq}{\end{eqnarray}}
\newcommand{\nn}{\nonumber}
\def\ltap{\ \raise.3ex\hbox{$<$\kern-.75em\lower1ex\hbox{$\sim$}}\ }
\def\gtap{\ \raise.3ex\hbox{$>$\kern-.75em\lower1ex\hbox{$\sim$}}\ }
\def\CO{{\cal O}}
\def\CL{{\cal L}}
\def\CM{{\cal M}}
\def\tr{{\rm\ Tr}}
\def\CO{{\cal O}}
\def\CL{{\cal L}}
\def\CM{{\cal M}}
\def\mpl{M_{\rm Pl}}
\newcommand{\bel}[1]{\be\label{#1}}
\def\al{\alpha}
\def\bt{\beta}
\def\eps{\epsilon}
\def\eg{{\it e.g.}}
\def\ie{{\it i.e.}}
\def\mn{{\mu\nu}}
\newcommand{\rep}[1]{{\bf #1}}
\def\be{\begin{equation}}
\def\ee{\end{equation}}
\def\bea{\begin{eqnarray}}
\def\eea{\end{eqnarray}}
\newcommand{\eref}[1]{(\ref{#1})}
\newcommand{\Eref}[1]{Eq.~(\ref{#1})}
\newcommand{\gsim}{ \mathop{}_{\textstyle \sim}^{\textstyle >} }
\newcommand{\lsim}{ \mathop{}_{\textstyle \sim}^{\textstyle <} }
\newcommand{\vev}[1]{ \left\langle {#1} \right\rangle }
\newcommand{\bra}[1]{ \langle {#1} | }
\newcommand{\ket}[1]{ | {#1} \rangle }
\newcommand{\ev}{{\rm eV}}
\newcommand{\kev}{{\rm keV}}
\newcommand{\Mev}{{\rm MeV}}
\newcommand{\gev}{{\rm GeV}}
\newcommand{\tev}{{\rm TeV}}
\newcommand{\mev}{{\rm MeV}}
\newcommand{\meV}{{\rm meV}}
\newcommand{\mnu}{\ensuremath{m_\nu}}
\newcommand{\nnu}{\ensuremath{n_\nu}}
\newcommand{\mlr}{\ensuremath{m_{lr}}}
\newcommand{\acc}{\ensuremath{{\cal A}}}
\newcommand{\mav}{MaVaNs}
\newcommand{\disc}[1]{{\bf #1}} 
\newcommand{\mh}{{m_h}}
\newcommand{\hb}{{\cal \bar H}}
\newcommand{\me}{\mbox{${\rm \not\! E}$}}
\newcommand{\met}{\mbox{${\rm \not\! E}_{\rm T}$}}
\newcommand{\MPl}{M_{\rm Pl}}
\newcommand{\ra}{\rightarrow}

\newcommand{\dbar}{\bar{D}}

\newcommand{\slashchar}[1]{\setbox0=\hbox{$#1$}   
   \dimen0=\wd0                                     
   \setbox1=\hbox{/} \dimen1=\wd1                   
   \ifdim\dimen0>\dimen1                            
      \rlap{\hbox to \dimen0{\hfil/\hfil}}          
      #1                                            
   \else                                            
      \rlap{\hbox to \dimen1{\hfil$#1$\hfil}}       
      /                                             
   \fi}                                             %

\pagestyle{plain}

\title{An Effective Theory of Dirac Dark Matter}

\author{Roni Harnik}
\affiliation{SITP, Physics Department, Stanford University, 
Stanford, CA 94305 and \\
SLAC, Stanford University, Menlo Park, CA 94025}
\author{Graham D. Kribs}
\affiliation{Department of Physics and 
Institute of Theoretical Science, \\
University of Oregon, Eugene, OR 97403}
\preprint{}
\date{\today}

\begin{abstract}

A stable Dirac fermion with four-fermion interactions to 
leptons suppressed by a scale $\Lambda \sim 1$~TeV 
is shown to provide a viable candidate for dark matter.
The thermal relic abundance matches cosmology, while nuclear 
recoil direct detection bounds are automatically avoided
in the absence of (large) couplings to quarks.
The annihilation cross section in the early Universe is the
same as the annihilation in our galactic neighborhood.
This allows Dirac fermion dark matter to naturally explain the
positron ratio excess observed by PAMELA with a minimal boost factor, 
given present astrophysical uncertainties.  
We use the Galprop program for propagation of signal and
background; we discuss in detail the uncertainties resulting 
from the propagation parameters and, more importantly, 
the injected spectra.
Fermi/GLAST has an opportunity to see a feature in the 
gamma-ray spectrum at the mass of the Dirac fermion.
The excess observed by ATIC/PPB-BETS may also be explained
with Dirac dark matter that is heavy.
A supersymmetric model with a Dirac bino provides a viable UV model 
of the effective theory. 
The dominance of the leptonic operators, and thus the observation 
of an excess in positrons and not in anti-protons, 
is naturally explained by the large  hypercharge and low mass 
of sleptons as compared with squarks.  Minimizing the boost factor 
implies the right-handed selectron is the lightest slepton, 
which is characteristic of our model.  Selectrons (or sleptons) 
with mass less than a few hundred GeV are an inescapable consequence 
awaiting discovery at the LHC\@.

\end{abstract}

\maketitle

\section{Introduction}

We propose a new dark matter candidate that is a stable Dirac 
fermion with four-fermion couplings to leptons.  
This candidate automatically avoids the direct detection
bounds from nuclear recoil direct detection experiments in 
the absence of couplings to quarks.  The annihilation rate
through a four-fermion leptonic operator yields a thermal
relic abundance consistent with cosmological data 
for an electroweak scale mass and TeV scale suppressed 
operator.  In this paper we present the effective theory, 
the application to the recent PAMELA positron ratio observations
\cite{Adriani:2008zr}
and present a supersymmetric realization.  

The annihilation rate of Dirac dark matter into leptons is 
not velocity suppressed,
similar to KK dark matter \cite{Cheng:2002ej,Hooper:2004xn,Hooper:2004bq},
allowing for an unsuppressed annihilation rate in our local
galactic neighborhood.
This provides a compelling explanation of the up-turn in the 
ratio of positron flux to electron plus positron flux 
observed by the PAMELA collaboration for positron energies above 
about 10 GeV \cite{Adriani:2008zr}.  Several groups have already considered 
various implications of this result using pre-publication 
\cite{Cirelli:2008id,%
Bergstrom:2008gr,%
Cirelli:2008jk,%
Barger:2008su,%
Cholis:2008hb,%
Cirelli:2008pk,%
Huh:2008vj,%
ArkaniHamed:2008qn,%
ArkaniHamed:2008qp,%
Finkbeiner:2008qu,%
Pospelov:2008jd,%
Hooper:2008kg,%
Hisano:2008ti,%
Yuksel:2008rf,%
Kamionkowski:2008gj,%
Khalil:2008ps,%
Serpico:2008te}
and post-publication 
\cite{Nelson:2008hj,%
Donato:2008jk,%
Cholis:2008xx,%
Nomura:2008xx%
} 
PAMELA data.
In our effective theory, we find that minimal enhancement in the 
local annihilation rate (a minimal boost factor) is required given 
the uncertainties in the local dark matter density as well as the 
present uncertainties in the background flux of electrons and positrons.  

We first present our effective theory of Dirac dark matter.
Our discussion is general; we consider all operators consistent 
with the symmetries.  We focus on the four-fermion leptonic 
dimension-6 operator that gives the largest annihilation rate
into positrons.
We then proceed to calculate the thermal relic density resulting
from these operators, determining the relative ranges of
parameters that are consistent with the cosmological abundance
$\Omega h^2 \simeq 0.1143 \pm 0.0034$ \cite{Komatsu:2008hk}.
Since the thermally-averaged annihilation rate is not
velocity suppressed, the same cross section that determines
the thermal relic abundance also determines the local annihilation
rate in the galaxy.  This lack of velocity dependence also
eliminates the uncertainty associated with the local WIMP
velocity distribution.

We then discuss the indirect detection signal of Dirac dark matter
through annihilation into leptons and anti-leptons. 
Understanding the size and uncertainties of background
sources of electrons and positrons from non-dark matter sources
is discussed in detail.  We do not consider point astrophysical
sources (such as pulsars \cite{Hooper:2008kg,Yuksel:2008rf}), 
and instead concentrate on secondary
production \cite{Moskalenko:1997gh}
from protons scattering off other protons, 
using the Galprop code \cite{galprop}.  Since the PAMELA collaboration
has not provided the absolute flux of electrons or 
positrons, we must
use other experimental results to determine the background spectra.  
We have analyzed the electron spectra of several recent
experiments, including 
AMS-01 \cite{Alcaraz:2000bf},
ATIC \cite{ATIC}, 
BETS \cite{Torii:2001aw,Torii:2008xu}, 
CAPRICE \cite{CAPRICE}, 
HEAT \cite{DuVernois:2001bb}, and 
MASS \cite{Grimani:2002yz}, 
and thus determined the best-fit per experiment and range of 
experimentally measured background electron flux.  
The positron injection spectrum and its spatial distribution 
is determined from Galprop as secondary production of positrons 
from primary cosmic rays (mostly protons) that scatter off other protons or nuclei.
This spectrum is normalized  by matching the well-measured proton flux,
which we have cross-checked against the preliminary 
proton spectra results of PAMELA \cite{pamelaSLAC}, 
and by the spatial distribution of interstellar gas.

The propagation and energy loss of the positrons is folded in
consistently using Galprop to determine the positron spectrum 
and positron flux ratio that would be expected
at PAMELA or AMS-02.  We find the flux ratio spectrum matches 
the PAMELA data \cite{Adriani:2008zr} 
with either a light Dirac fermion that annihilates 
into $e^+e^-$, 
or democratically into all charged leptons $\ell^+\ell^-$,
$\ell = e,\mu,\tau$. 
Interpolations between these cases are
also consistent with the data.  The boost factor
is minimized for the lightest mass dark matter particle and with 
annihilation purely into $e^+e^-$.
We find it can be as small as $1$, for the case of 
a 100 GeV Dirac fermion annihilating only into $e^+e^-$, assuming 
the local \emph{average} relic density $\rho_{8.5} = 0.7$ GeV/cm$^3$ and an 
$e^-$ cosmic spectrum $\Phi_{e^-}(E)$ that falls as $E^{-3.15}$, \emph{or} for
$\rho_{8.5} = 0.3$ GeV/cm$^3$ and an $e^-$ spectrum that
falls as $E^{-3.5}$.  For canonical values we use in this paper, 
$\rho_{8.5} = 0.3$ GeV/cm$^3$ and $\Phi_{e^-}(E) \propto E^{-3.15}$, 
the boost factor is $5$.
Larger masses are permitted to the extent that a larger boost 
factor $B$ is plausible, scaling approximately as $B \propto M^2$.

Dirac dark matter with a minimal boost factor immediately implies 
that PAMELA should see a sharp drop in their positron spectrum 
above the mass of the Dirac fermion.  At exactly what mass we expect 
this drop to occur 
is nontrivially convoluted with the astrophysical uncertainties
of the background flux and the local dark matter density.
We will illustrate these uncertainties
and their impact on a Dirac fermion dark matter interpretation.
Other experiments, particularly Fermi/GLAST, should 
see a photon feature at the mass of the Dirac fermion, which results 
from final-state radiation off the charged lepton \cite{Birkedal:2005ep}.
This observation may occur before, simultaneous with, or after 
PAMELA provides their spectral data up to their experimental limit.

There have also been hints of an excess in the flux of electrons
\emph{plus} positrons, reported by ATIC \cite{ATIC} 
for energies between about 300-500 GeV and 
PPB-BETS \cite{Torii:2008xu} for energies between about
200-500 GeV\@.  An earlier study of the electron flux 
with an emulsion chamber balloon experiment \cite{Kobayashi:1999he}
in the same energy range, however, did not observe an excess 
(see also \cite{Torii:2008xu}).
Nevertheless, this hint has been recently investigated and
interpreted in \cite{ArkaniHamed:2008qn,Cirelli:2008pk,Nelson:2008hj}.
If this hint persists and is confirmed by future data, 
it is straightforward to explain with Dirac dark matter
so long as the mass of the Dirac fermion is large.
Within the uncertainties in the electron spectra,
the boost factor needed to explain the PAMELA excess 
as well as a feature in the spectra observed by
ATIC/PPB-BETS can be as small as $16$ for $M = 400$ GeV\@.

Dirac fermion dark matter may interact in other ways, 
in particular with quarks.  However, vector interactions 
between a Dirac fermion and quarks are highly constrained by
the absence of direct detection through nuclear recoil,
a fact well-known from neutrino dark matter
(for example, see \cite{Belanger:2007dx}).
Other implications, such as indirect detection through
accumulation and annihilation in the Sun are certainly
possible if the dark matter scattering off protons in the Sun 
is efficient enough to bring capture and annihilation 
into equilibrium.  For example, if the dominant annihilation mode 
were into \emph{left-handed} leptons, then a neutrino signal 
becomes a distinct possibility.

Finally, we show that a viable ultraviolet (UV) completion for 
a Dirac fermion that automatically has the desired properties to explain
the PAMELA positron excess is a Dirac bino in supersymmetry.  
Dirac gauginos occur
automatically in supersymmetric models with an exact 
$R$-symmetry \cite{Kribs:2007ac}.  Model-building the
lightest supersymmetric particle to be a Dirac bino
is a fairly straightforward variation of the model proposed 
in \cite{Kribs:2007ac} (albeit with some supergravity 
subtleties that we outline below).
Four-fermion operators arise from the exchange of sfermions,
and thus the relative strength of different operators can
be directly interpreted in terms of the strength of couplings
and the relative hierarchy of sfermion masses.
Since a Dirac bino couples through hypercharge, the largest
four-fermion couplings are to the right-handed sleptons.
Moreover, models of supersymmetry run from higher scales
often give the smallest masses to the right-handed sleptons.
This provides a compelling explanation of the dominance
of right-handed leptonic operators for annihilation.
Minimizing the boost factor implies the right-handed selectron 
is the lightest slepton, which is characteristic
of our model.  
We can use the annihilation rate to predict the range of 
slepton masses, less than a few hundred GeV, which provides
fantastic opportunities for discovery and further study 
at the LHC\@.

\section{Dirac Dark Matter}

An effective field theory of Dirac dark matter is extremely
simple.  The purported dark matter particle is a Dirac fermion 
$D$ that transforms under an exact continuous global symmetry $U(1)_D$,
$D \rightarrow D e^{i \theta_D}$.  For the purposes of this discussion, 
we assume the full $U(1)_D$ global fermion number to be conserved,
but a discrete subgroup may also suffice to ensure the
particle is stable.  
The global $U(1)_D$ allows a vector-like mass term for $D$.
The Lagrangian for this electroweak- and color-neutral particle is
\begin{eqnarray}
{\cal L} &=& i \dbar \slashchar{\partial} D - M \dbar D
\end{eqnarray}

A general effective field theory for $D$ interacting with the SM
can be written as
\begin{eqnarray}
{\cal L} = \sum_{n,i} \frac{{\cal O}_i^{(n)}}{\Lambda^{n-4}} \; ,
\end{eqnarray}
where the leading interactions are through $n$-dimensional
operators labeled by $i$.  The only operator at dimension-5,
\begin{eqnarray}
{\cal O}^{(5)} &=& 
\frac{\dbar D H^\dagger H}{\Lambda} 
\label{dim-5}
\end{eqnarray}
while there are many operators at dimension-6, including
\begin{eqnarray}
{\cal O}^{(6)}_{f_L} &=& c^L_f \frac{\dbar \gamma^\mu D 
                               \overline{f} \gamma_\mu P_L f}{\Lambda^2} 
\label{left-op} \\
{\cal O}^{(6)}_{f_R} &=& c^R_f \frac{\dbar \gamma^\mu D 
                               \overline{f} \gamma_\mu P_R f}{\Lambda^2}
\label{right-op}
\label{dim-6}
\end{eqnarray}
We did not write interactions involving $\dbar \gamma^\mu \gamma^5 D$
because they lead to either velocity-suppressed or mass-suppressed
interactions.
Since left-handed SM fields transform as doublets under $SU(2)_L$, 
for leptons $c^L_\ell$ automatically leads to equal interaction strength 
to charged left-handed leptons and their neutrino partners.  
Similarly, $c^R_\ell$ leads to an interaction with just right-handed 
charged leptons.  The right-handed four-fermion leptonic operator will 
be the main focus of this paper.

Before proceeding, it is worthwhile to examine effects 
of the other operators.  The leading dimension-5 operator, 
after electroweak symmetry breaking, leads to three effects:
a coupling to a pair of Higgses, a coupling to just one Higgs, 
and a shift in the Dirac mass.
The coupling $\dbar D h^2$ allows dark matter to 
annihilate into a pair of Higgses.  The single Higgs coupling,
$(v/\Lambda) \times \dbar D h$ has two interesting
effects:  If $M < m_h/2$, the Higgs could decay into a pair of 
dark matter fermions,
with a branching ratio proportional to $(v/\Lambda)^2$ that could 
compete with $h \rightarrow \overline{b}b$ if $m_h \lsim 2 M_W$.
If $M > m_h/2$, Higgs decays are essentially unaffected by the 
presence of Dirac dark matter.  Nevertheless, this effective
interaction can lead to dark matter annihilation through an 
$s$-channel Higgs.  
It is a model-dependent question whether the dimension-5 Higgs
interaction is important or relevant.  (For a Dirac bino,
it is irrelevant, as we will see below.)

The dimension-6 operators include interactions between
Dirac dark matter with both quarks and leptons, 
left-handed and right-handed.  Vector-like interactions
between dark matter and quarks leads to a spin-independent
scattering cross section that is strongly constrained by
direct dark matter searches such as CDMS \cite{Ahmed:2008eu}
and Xenon \cite{Angle:2007uj}.  Since no
direct detection signal has been (unambiguously) observed,
and PAMELA has not found an excess in anti-protons \cite{Adriani:2008zq},
we will not consider the four-fermion quark operators.

This leaves the four-fermion lepton operators.
These operators are unconstrained by direct detection
experiments because cross sections of dark matter with
atomic electrons are suppressed by a tiny form factor 
\cite{Bernabei:2007gr}.
The PAMELA positron ratio excess suggests maximizing the 
annihilation rate into positrons through the single operator 
$O^{(6)}_{e_R}$.  We will also consider what happens in the
right-handed flavor-democratic case with all three operators 
$O^{(6)}_{\ell_R}$ are present with equal strength, as well as 
mention what happens when both left-handed and right-handed 
flavor-democratic operators $O^{(6)}_{\ell_L},O^{(6)}_{\ell_R},$ 
are present.

Dirac dark matter coupling to left-handed leptons provides an 
interesting possibility, since annihilation necessarily 
also yields neutrinos.  A combination of small interactions 
with quarks as well as left-handed leptons may yield an interesting 
indirect signal resulting from the annihilation of Dirac dark matter 
in Sun.  We leave this interesting calculation for future work.

\section{Relic Abundance}

The relic abundance of a Dirac fermion has been calculated in
\cite{Srednicki:1988ce,Hsieh:2007wq}.  Unlike Majorana
particles \cite{Goldberg:1983nd}, the leading order contribution 
to the annihilation directly into leptons is \emph{not} 
velocity-suppressed.
This proves extremely convenient in providing a model-independent 
relationship between the thermal relic density at freeze-out in
the early universe and the galactic annihilation rate occurring today.
This relationship provides a tight constraint on the size and shape 
of the expected positron flux, making Dirac dark matter highly predictive.

In the presence of $O^{(6)}$, the thermally-averaged 
annihilation cross section can be written quite generally as
\begin{eqnarray}
\langle \sigma_{\dbar D} v \rangle &=& \frac{M^2}{2 \pi} \sum_f
\frac{|c^{L(R)}_f|^2}{\Lambda^4}
\label{annih-eq}
\end{eqnarray}
where the sum is over all dimension-6 four-fermion operators.
We have neglected the higher order temperature-dependent
corrections proportional to $1/x_F \equiv T/M$, which 
shift the cross section by less than 10\%.  
We have also ignored the masses of the final state fermions,
which is of course an excellent approximation for leptons.
All of the model-dependence is buried in the couplings $c_f$. 
Let's focus first on the case where the only open annihilation
channel is to right-handed leptons, i.e., $c^R_{e} = 1$
and all other $c$'s vanish.

The thermal relic abundance for Dirac fermions results in an
equal abundance of particle $D$ and anti-particle $\dbar$ 
(since we are assuming no pre-existing asymmetry in
$D$ number).  Consequently, the relevant cross section that
enters both the thermal relic abundance as well as the annihilation
rate in the galaxy is 
\cite{Srednicki:1988ce}
\begin{eqnarray}
\sigma_{\rm eff} &=& \sum_{ij} \frac{n_i n_j}{n^2_{\rm tot}} 
\langle \sigma_{ij} v \rangle \\
                 &=& \frac{1}{2} \langle \sigma_{\dbar D} v \rangle \; .
\end{eqnarray}
The factor of $1/2$ accounts for only two of four annihilation rates 
($D\bar{D}$ and $\bar{D}D$ but not $DD$ or $\bar{D}\bar{D}$) 
being nonzero.  The thermal relic abundance is then
\begin{eqnarray}
\Omega h^2 = x_F \frac{8.54 \times 10^{-11} \; {\rm GeV}^{-2}}{
\sqrt{g_\star} \langle \sigma_{\dbar D} v \rangle/2} \; ,
\end{eqnarray}
where $g_\star \simeq 96$ is the number of relativistic degrees
of freedom at freeze-out.  

\begin{figure}
\begin{center}
\includegraphics[width=0.45\textwidth]{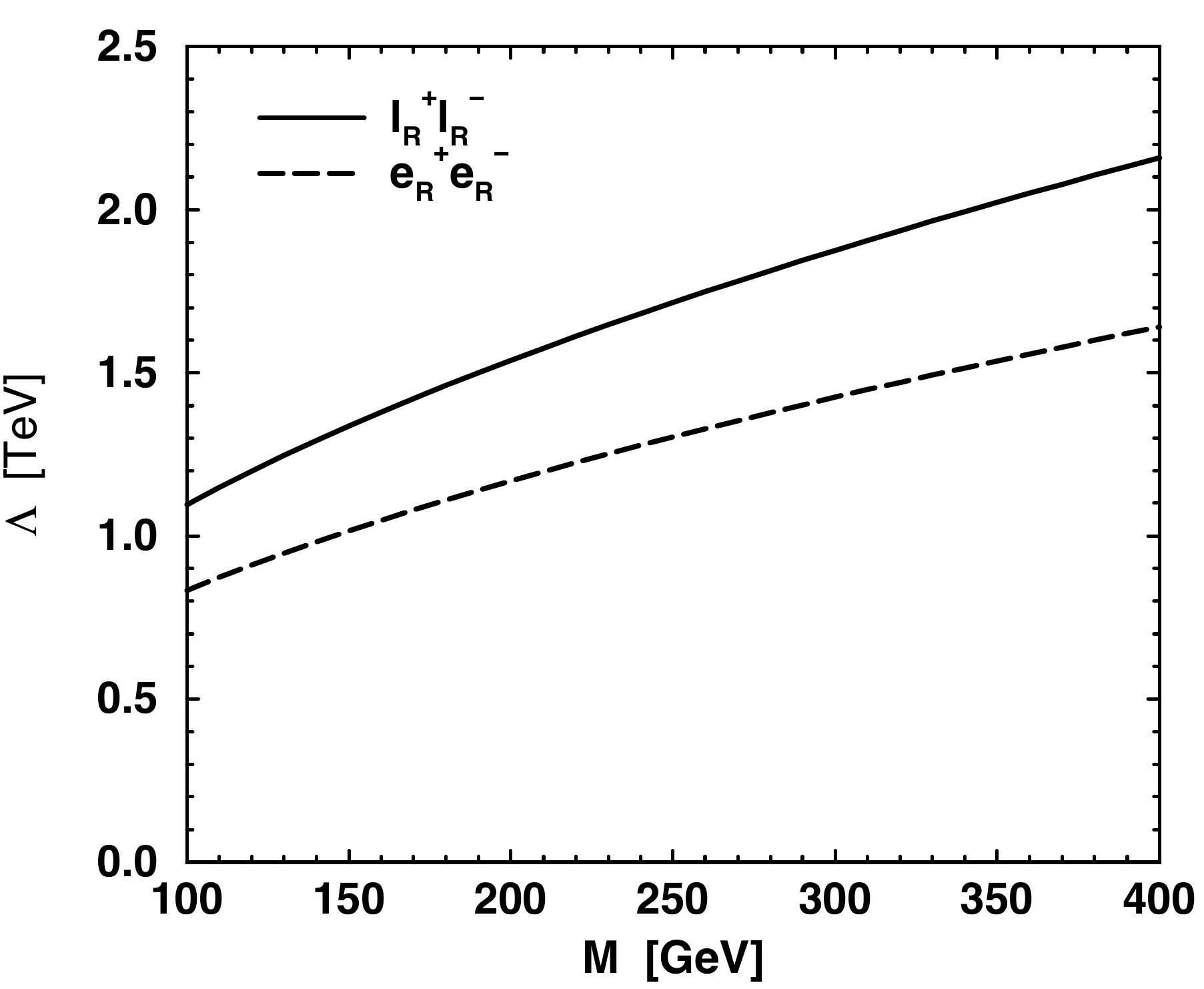}
\end{center}
\caption{Cutoff scale $\Lambda$ as a function of the 
Dirac dark matter fermion mass $M$
that gives the thermal relic abundance $\Omega h^2 = 0.114$,
consistent with cosmological data.
The top curve corresponds to the flavor-democratic
scenario, $c_{e_R} = c_{\mu_R} = c_{\tau_R} = 1$,
while the lower curve corresponds to electrons only
$c_{e_R} = 1$.
In both cases we took only right-handed leptons 
for simplicity; adding left-handed leptons is trivial.}
\label{effomega-fig}
\end{figure}

Using cosmological data to fix the thermal relic abundance 
to be $\Omega h^2 = 0.114$, we can determine the leading order
(velocity-independent part) annihilation cross section,
\begin{eqnarray}
\langle \sigma_{\dbar D} v \rangle &=& (1.25 \; {\rm pb}) \frac{x_F}{21} 
\sqrt{\frac{96}{g_\star}}
\label{cross-section-eq}
\end{eqnarray}
In Fig.~\ref{effomega-fig} we show
the relationship between $\Lambda$ and $M$ to obtain the
thermal relic abundance consistent with cosmological data.
The range of masses shown is illustrative.  A lower bound on $M$ 
can be established from the absence of a single photon plus missing
energy signal at LEPI that would occur with the dimension-6
operator combined with an initial state photon.  
By contrast, LEPII does not place strong bounds on this process 
(for example, see \cite{Ambrosanio:1995it}), essentially because 
the cross section is suppressed by $\alpha_{\rm em}$ and phase space 
that causes the signal to be too small to be seen above background.  
This suggests $M$ could be as low as about 50 GeV\@.  But as we will see, 
to explain the PAMELA positron ratio excess we need $M \lsim 100$ GeV,
and thus there is no direct limit from LEPII\@.

\section{Positrons from Annihilation of Dirac Dark Matter}

\subsection{Backgrounds and Galactic Propagation}

Determining the background electron and positron flux is of
utmost importance to establish that the positron ratio excess
does, in fact, exist.  The most complete calculation of the background
fluxes of cosmic rays comes from the Galprop code \cite{galprop},
where antimatter is generated as secondary production
from protons scattering off other protons and lighter nuclei.
We will briefly explain the inputs to the code, the various
propagation model and parameter dependencies, and thus 
our estimates of the uncertainties in the background.
We use Galprop to propagate both signal and background.  
This is the only consistent way to treat propagation uncertainties.  
We have, nevertheless, cross-checked our signal using semi-analytic 
treatments of propagation \cite{Baltz:1998xv}.

Galprop is, for cosmic rays, similar in spirit to Pythia 
for collider experiments.  Just as Pythia incorporates theoretical
calculations, such as cross sections, as well as semi-analytic 
techniques, such parton showering,
Galprop also incorporates both theoretical and experimentally-driven
models and assumptions to predict cosmic ray spectra.
There are three inputs to the code important for our analysis:
\begin{itemize}
\item[1.] The electron source spectrum.
\item[2.] The nuclei source spectrum.
\item[3.] The propagation model and associated parameters.
\end{itemize}
Other important inputs include nuclear cross sections,
interstellar gas distribution, etc \cite{galprop}.

The origin of the high-energy background spectrum of nuclei and 
electrons in the galaxy is presumed to come from supernovae,  
though it is at present not well understood.
Galprop does not attempt to determine these spectra
from first principles.  Instead, the spectra are assumed
to arise from an ``injected'' power-law input flux 
with coefficients, breaks, spatial distribution, and normalization 
determined by fitting to astrophysical data.  Galprop self-consistently 
``propagates'' all of the cosmic rays within  
galactic magnetic fields, allowing for particle collisions that
result in secondary production of antiprotons, positrons,
as well as secondary production of electrons, protons, etc.

The spectra in interstellar space differs from observations 
near Earth due to the solar modulation effect arising from the 
solar wind.  This is expected to shift the observed energy
by of order $0.6$ GeV \cite{Casolino:2008zm}.
We focus only on the 
data above 5 GeV, thereby minimizing this systematic error.

Since PAMELA has not yet provided the absolute fluxes of 
electrons or positrons, we are forced to use data from
other experiments to determine the absolute background flux.
AMS-01 \cite{Alcaraz:2000bf},
ATIC \cite{ATIC}, 
BETS \cite{Torii:2001aw,Torii:2008xu}, 
CAPRICE \cite{CAPRICE}, 
HEAT \cite{DuVernois:2001bb}, and
MASS \cite{Grimani:2002yz}, have measured the
electron flux, with or without charge identification.
We have performed a weighted least-squares fit to their data
for energies larger than 5 GeV\@.  For BETS and MASS we used
their reported their best fit, since their energy range began
above 5 GeV\@.
Our results are given in Table~\ref{electron-data-table}.
\begin{table}
\begin{center}
\begin{tabular}{r|c}
Experiment & power law index $\alpha$ \\ \hline
AMS-01 \cite{Alcaraz:2000bf} & $3.15 \pm 0.04$ \\
ATIC \cite{ATIC}             & $3.14 \pm 0.08$ \\
BETS \cite{Torii:2001aw,Torii:2008xu} & $3.05 \pm 0.05$ \\
CAPRICE \cite{CAPRICE}       & $3.47 \pm 0.34$ \\
HEAT \cite{DuVernois:2001bb} & $2.82 \pm 0.16$ \\ 
MASS \cite{Grimani:2002yz}   & $2.89 \pm 0.10$
\end{tabular}
\end{center}
\caption{Our weighted least-squares best fit to the
electron flux, $\Phi_{e^-}(E) \propto E^{-\alpha}$, measured by the 
various experiments.  The BETS best fit was taken from \cite{Torii:2008xu};
their best fit to just the lower energy data 
between 10 to 100 GeV is $3.00 \pm 0.09$ \cite{Torii:2001aw};
the error is assumed to be $1\sigma$.
The MASS best fit taken from \cite{Grimani:2002yz};
the error is assumed to be $1\sigma$.
We emphasize that our reported errors for the other experiments
are purely statistical (95\% CL) with regard to fitting data 
(with errors) to a power law, and do not necessarily reflect 
the individual experiments' precision.}
\label{electron-data-table}
\end{table}
A very conservative interpretation of the data is that the 
observed electron flux is falling as $E^{-3.15 \pm 0.35}$
for $E > 5$ GeV, which spans all of the central best-fit values 
of the experiments.  Another approach to the uncertainties 
in the electron spectra can be found in \cite{Casadei:2004sb}.
Their result for the electron flux is that it falls as
$E^{-3.44 \pm 0.03}$, which is within our range,
though with what seems to us to be an unrealistically 
small error.

The spectra of positrons is determined from secondary
production, after protons (or heavier nuclei) inelastically 
collide into other protons or nuclei, emitting charged pions
that decay into positrons.  This requires simulating
networks of hadron interactions and decays, using nuclear
and particle physics data.
The positron flux is thus ultimately determined by the 
injected nucleon spectrum, nuclear cross sections and the propagation model 
and parameters.

By fitting the resulting nucleon spectra to data, 
the injected nucleon spectrum and propagation parameters
can be well constrained.  A recent study by \cite{Lionetto:2005jd}
used Galprop to fit to the proton spectra, the $B/C$ ratio, 
and other data to determine the best-fit and a range of
propagation parameters.  We use their results in determining 
the propagation model and parameters that best reproduce
the nucleon spectra.  Their study \cite{Lionetto:2005jd}
considered propagation with convection (``DC'' model), 
with reacceleration (``DR'' model), and reacceleration with 
a break in the spectra (``DRB'' model).  They also considered a 
``min'', ``max'', and ``best'' set of propagation parameters
for each model.  We found that using the default proton 
injection spectrum in Galprop, combined with either the
``min'' or ``max'' sets of propagation parameters,
generally gave a considerably worse fit to the experimentally 
observed proton spectrum \cite{Amsler:2008zz,pamelaSLAC}.
Since positrons derive from protons, we opted to consider 
only their ``best'' fits.  We should emphasize that these three models 
do not represent the full uncertainty in propagation, but are 
rather meant to gain a quantitative understanding of the different 
spectra possible with qualitatively different models of propagation.  
Further studies of propagation effects can be found in 
\cite{Delahaye:2008ua}.
In the end, the propagation parameter dependence is 
considerably milder than the present uncertainty arising 
from the background electron spectrum.

\begin{figure}
\begin{center}
\hspace*{-0.04\textwidth}
\includegraphics[width=0.55\textwidth]{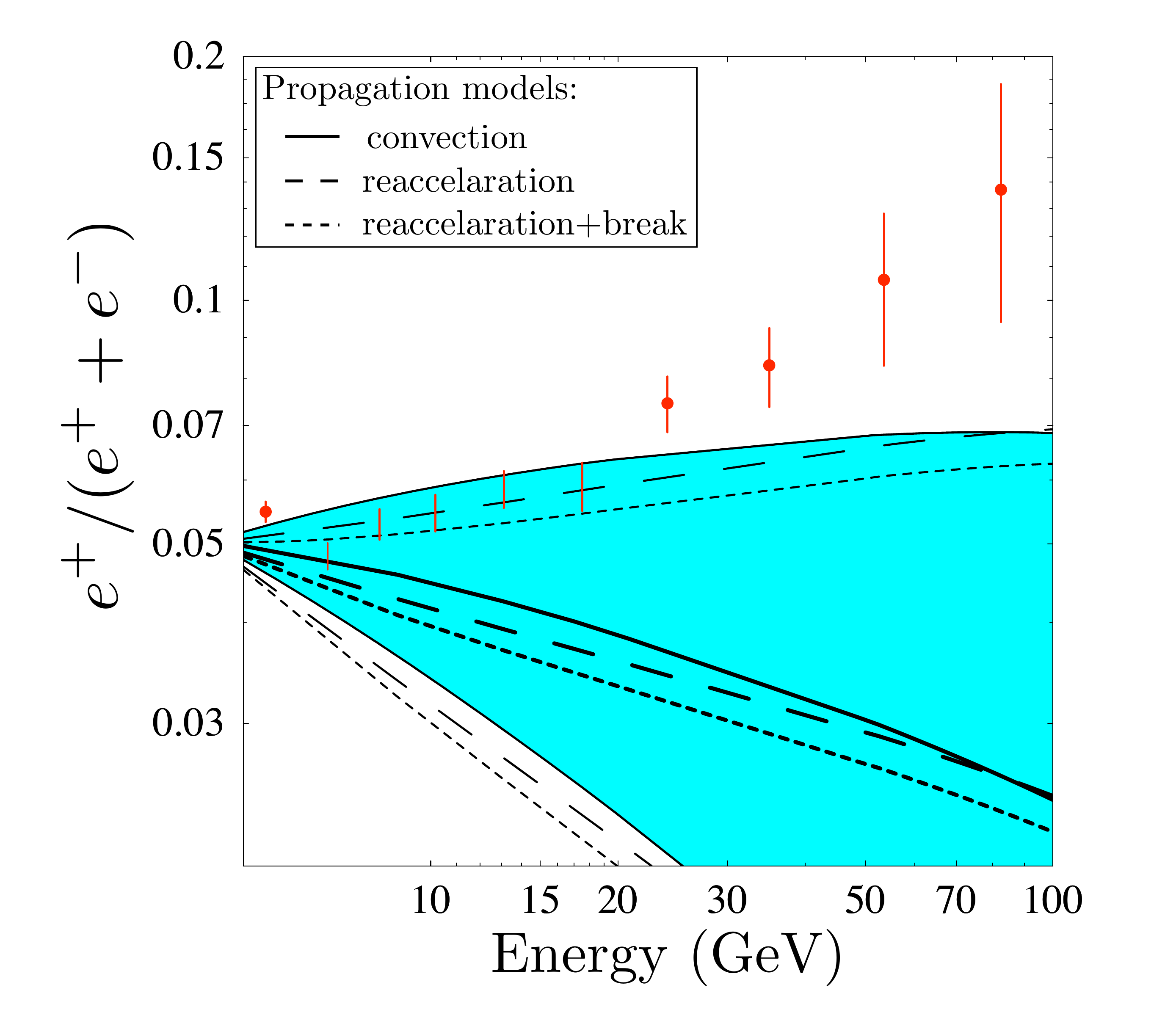}
\vspace*{-9mm}
\end{center}
\caption{The positron ratio assuming background only as calculated 
by Galprop for the 3 propagation models described in the text, 
DC (solid), DR (long dashed) and DRB(short dashed).  The central 
thick lines assume an electron spectral spectrum 
$\Phi_{e^-}(E) \propto E^{-3.15}$ whereas 
the thinner lines above and below show the affect of varying the 
electron spectrum by 
$\Phi_{e^-}(E) \propto E^{-3.5}$ and $E^{-2.8}$, respectively,
within the range as determined by
Table~\ref{electron-data-table}.  The data is taken from the 
recent PAMELA observations \cite{Adriani:2008zr}.}
\label{fig-background}
\end{figure}

We therefore determined the background spectrum in the
following way.  Given a propagation model, the absolute
positron spectrum is determined.  Using the published 
PAMELA flux ratio data point at $4.5$ GeV \cite{Adriani:2008zr},
we inverted the positron flux to obtain the absolute electron flux 
$\Phi_{e^-}(4.5 \; {\rm GeV} = 2.5 \times 10^{-4}$ 
cm$^{-2}$~s$^{-1}$~sr$^{-1}$~GeV$^{-1}$,
which is our normalization for the background.
We then used our power-law best fit range to the electron 
data given in Table~\ref{electron-data-table} as the 
background electron flux.  This procedure assumes that any 
new physics contribution to the positron (or electron) flux 
at $4.5$ GeV is negligible (which we verify, \textit{ex post facto}, 
below). The resulting background positron fraction is shown in 
Figure~\ref{fig-background} for the three propagation models 
we have chosen (Thick lines). The uncertainties due to the variations 
in the electron spectral slope are also shown (thin lines and blue band 
for the DC model). As advertised earlier, the uncertainty in the 
electron spectrum currently dominates, and will hopefully lessen 
as the absolute electron flux from PAMELA, Fermi/GLAST, and other 
instruments are released.  However, despite the large uncertainty 
in background, the PAMELA shape and size, particularly at the 
highest energies, lies well outside our generous range of the 
predicted background from secondary production.  We will now explore 
the possibility that this excess can be explained by annihilation 
of Dirac dark matter particles.

\subsection{Positron Signal}

The same processes that freeze out a thermal relic abundance of 
Dirac dark matter also leads to an annihilation rate in our
galactic neighborhood.  Since the thermally-averaged annihilation
rate was dominated by the zero temperature limit, the same
annihilation rate, Eq.~(\ref{annih-eq}), also applies to
the annihilation happening in the galaxy today.  
This provides a model-independent relationship between 
annihilation rates, and provides one of the strongest 
constraints on a Dirac dark matter interpretation of the 
PAMELA excess.

The abundance in the local galactic neighborhood is 
typically taken to be $\rho_{8.5} = 0.3 \; {\rm GeV}/{\rm cm}^3$
\cite{Amsler:2008zz}.
We assume an isothermal halo profile, where
\begin{eqnarray}
\rho(r) = \rho_{8.5} \frac{r_{8.5}^2 + a^2}{r^2 + a^2}
\end{eqnarray}
with $a = 5$ kpc.  Our results are not strongly sensitive 
to the choice of profile,
since most energetic positrons arrive from our galactic
neighborhood, of order $1$~kpc, where the dark matter density
is not nearly as uncertain as it is in the galactic center.
The precise local \emph{average} dark matter density is itself 
subject to uncertainties.  Since this is a simple scaling of the 
signal, we will fold this uncertainty into the boost factor.
But of course it should be remembered that, for example,
a boost factor of $4$ could be equivalently obtained by
scaling $\rho_{8.5}$ up by a factor of 2, which is within
the uncertainties \cite{Bergstrom:1997fj,Amsler:2008zz}.

In addition to annihilation within the smooth dark matter halo, 
it has been suggested that indirect signals of dark matter annihilation 
could be boosted due to a large degree of clumpiness in our halo.
Such clumps of dark matter may be a remnant of the hierarchical 
build-up of galactic halos from small to large
(e.g.\ \cite{Diemand:2005vz}).
In particular, if the Earth happens to be near a dense dark matter clump, 
annihilation signals may be enhanced, though this does seem to be 
a probable scenario.  Recent many body simulations show that though 
a boost factor of order a few is possible, while a boost exceeding
of order 20 in the positron signal appears unlikely \cite{Lavalle:1900wn}.

The basic physics that leads to a positron flux from 
dark matter annihilation is twofold:  First, dark matter
annihilates into SM matter.  The annihilation could proceed
directly into $e^+e^-$, or into for example $\mu^+\mu^-$,
which then decays into electrons and positrons.  
Earlier analyses with pre-publication PAMELA data (e.g.\ \cite{Cholis:2008hb}) 
suggest that the annihilation channels $W^+W^-$, $b\bar{b}$, $q\bar{q}$ 
are not nearly as favorable as directly into $e^+e^-$ or $\ell^+\ell^-$,
given a velocity-independent annihilation cross section and 
minimizing boost factors.  We used DarkSUSY \cite{Gondolo:2005we}
to obtain the (at-source) energy distributions of positrons from 
annihilation into muons and taus.

The second component of a positron signal is the propagation
of a positron with a given energy from where it was created
to Earth. We propagate the signal positrons using Galprop 
for the three propagation models described above in the previous 
subsection.

\begin{figure}
\begin{center}
\hspace*{-0.04\textwidth}
\includegraphics[width=0.55\textwidth]{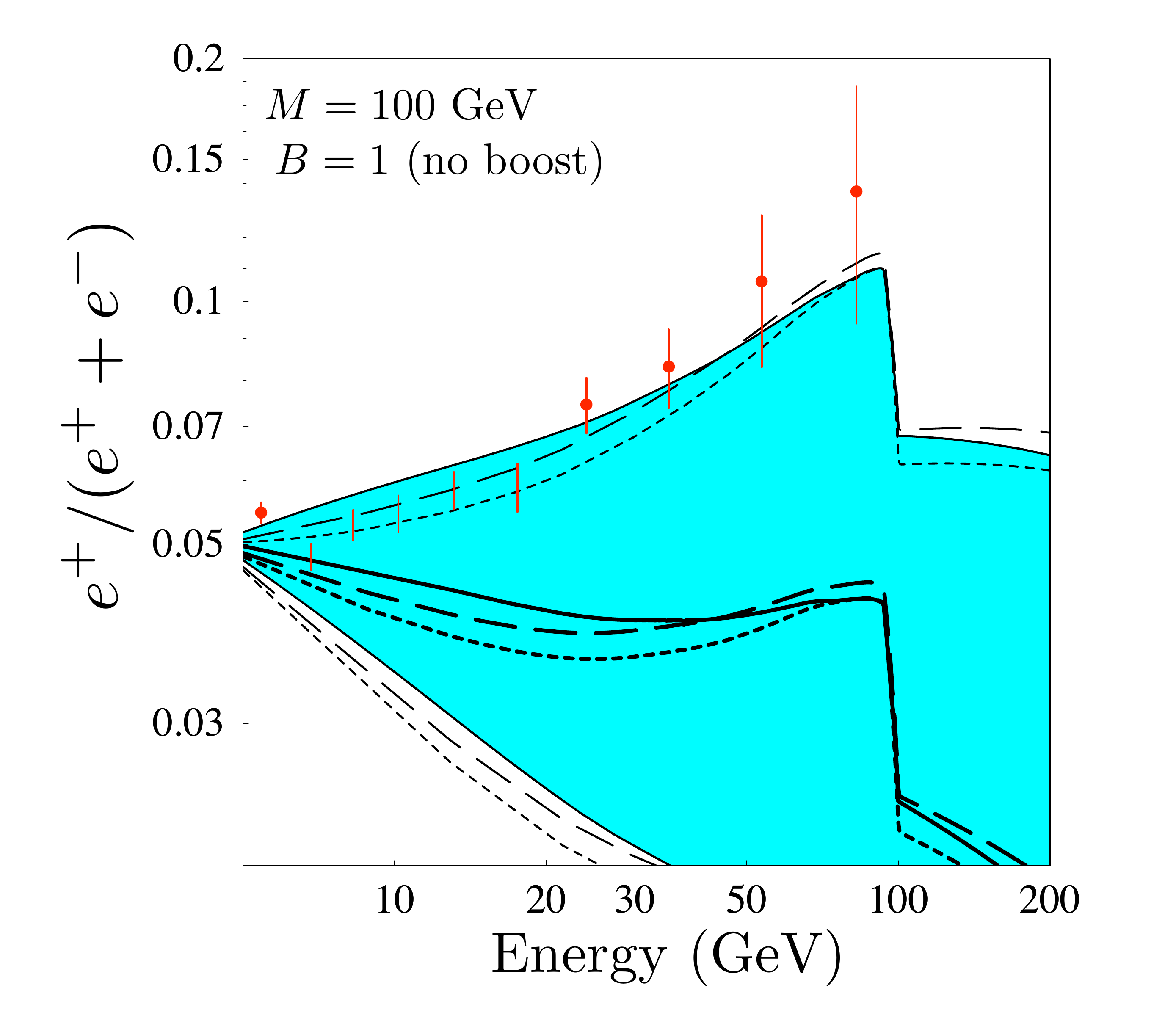}
\vspace*{-9mm}
\end{center}
\caption{The positron fraction from a 100 GeV 
Dirac dark matter particle that annihilates to right handed electrons. 
Three propagation models are plotted:  DC (solid), DR (long dash), 
and DRB (short dash), as well as the uncertainty due to variation of 
the electron spectral slope.  No boost factor was employed for 
this figure.  Within the present astrophysical uncertainties, 
the PAMELA data can be explained so long as the electron spectrum 
is quite steep, $\Phi_{e^-} \propto E^{-3.5}$, corresponding to the
top of the shaded blue band.}
\label{e100noboost}
\end{figure}

Our results are shown in a series of figures.
We begin with a Dirac dark matter candidate that couples only to 
right-handed electrons.  This benchmark model maximizes the signal. 
Indeed, as can be seen in Figure~\ref{e100noboost}, the PAMELA data 
lie within the uncertainty band of the expected signal, though 
fitting the data would require a rather steep electron spectrum, 
a hypothesis that will be surely be tested by PAMELA itself 
as well as Fermi/GLAST\@.
It should be stressed that in Figure~\ref{e100noboost} we use an 
annihilation cross section given by Eq.~(\ref{cross-section-eq})
which matches the relic abundance calculation.  Within the present
astrophysical uncertainties, we find no boost factor is required 
to explain the preliminary data. 
The case where Dirac dark matter annihilates to left-handed electrons 
is a simple halving of this signal since half of the annihilations 
are to neutrinos.

\begin{figure}
\begin{center}
\hspace*{-0.04\textwidth}
\includegraphics[width=0.55\textwidth]{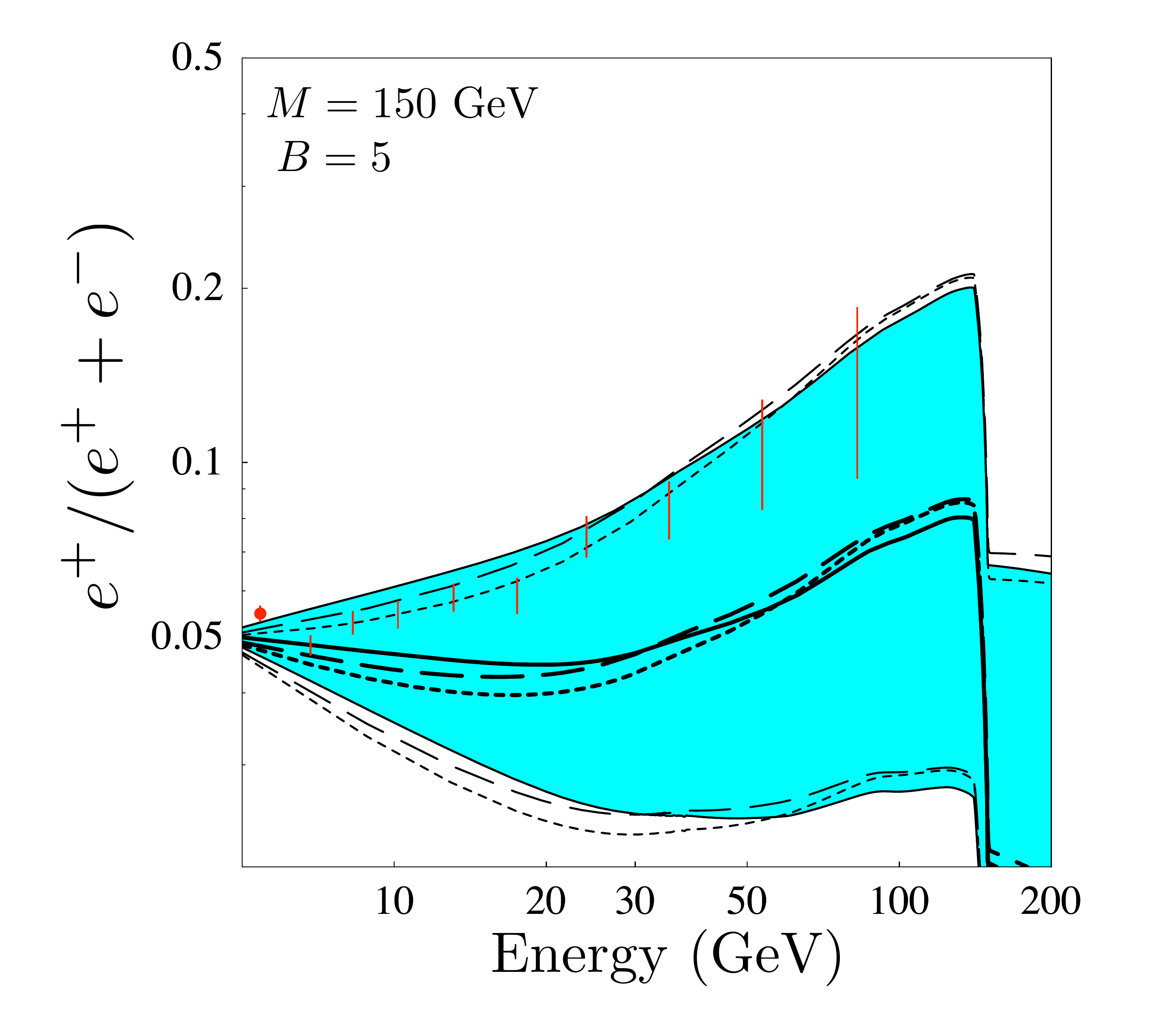}
\vspace*{-9mm}
\end{center}
\caption{Same as Fig.~\ref{e100noboost}, except for
$M = 150$ GeV and a boost factor of $5$.}
\label{e150boost5}
\end{figure}

\begin{figure}
\begin{center}
\hspace*{-0.04\textwidth}
\includegraphics[width=0.55\textwidth]{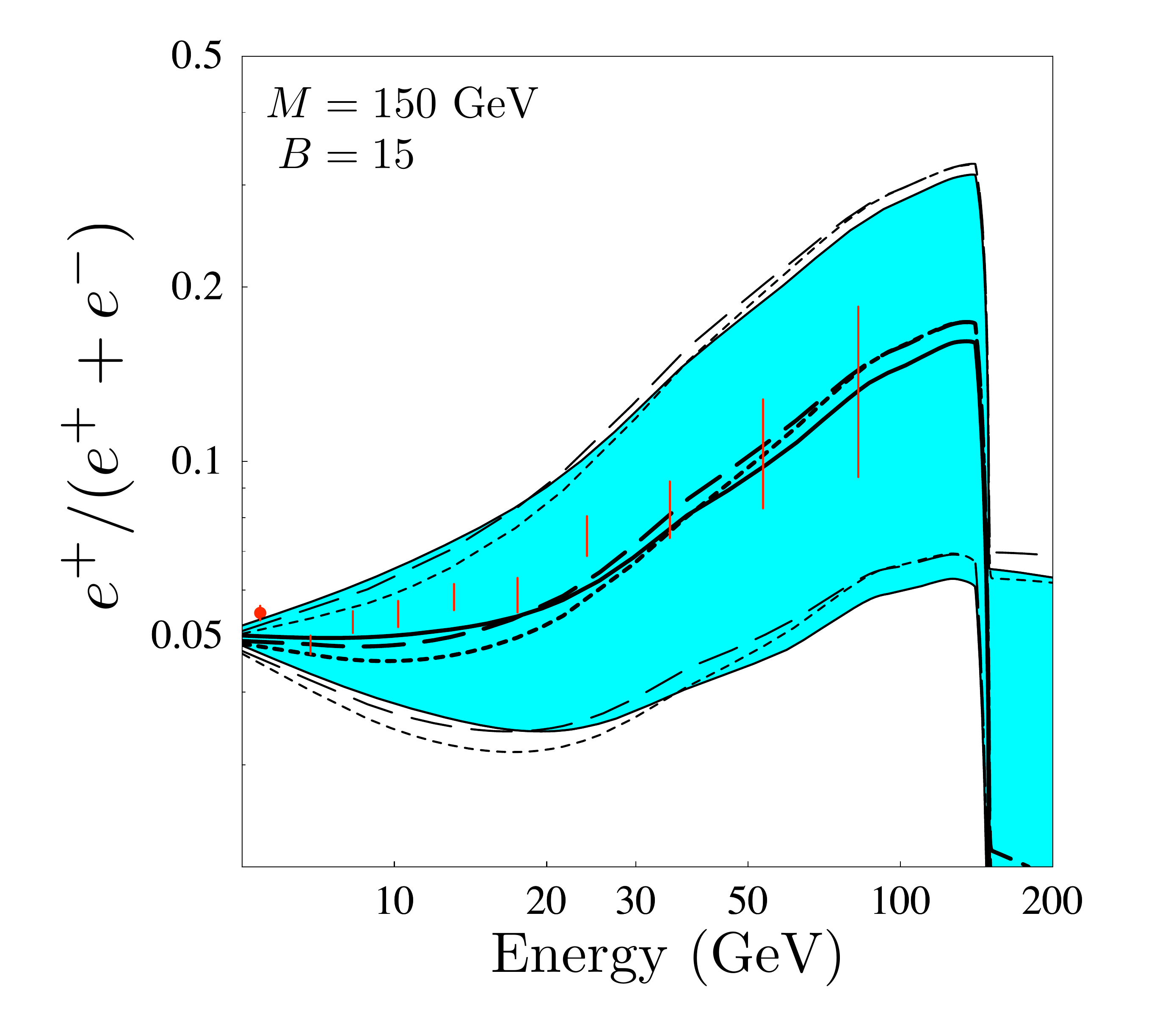}
\vspace*{-9mm}
\end{center}
\caption{Same as Fig.~\ref{e150boost5}, except with
a boost factor of $15$.}
\label{e150boost15}
\end{figure}

\begin{figure}
\begin{center}
\hspace*{-0.04\textwidth}
\includegraphics[width=0.55\textwidth]{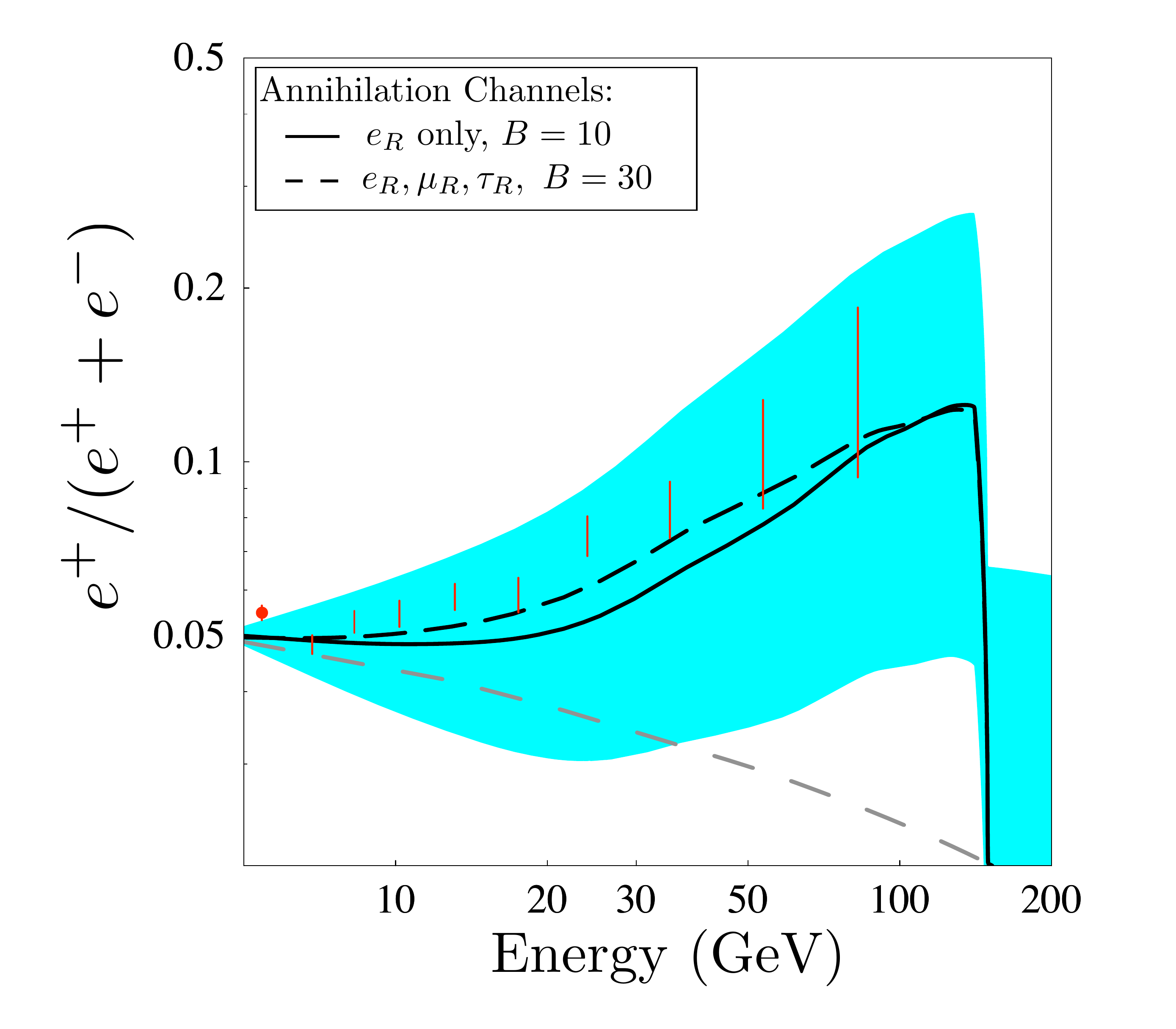}
\vspace*{-9mm}
\end{center}
\caption{The positron fraction from a 150 GeV 
Dirac dark matter particle that annihilates to leptons
assuming the DC propagation model.  The solid line
corresponds to annihilations to just right-handed electrons
with boost factor of $10$, while the dashed line 
corresponds to annihilations to all right-handed leptons
with boost factor of $30$.  The shaded blue band is the
same as previous figures.}
\label{DC150-emutau}
\end{figure}

\begin{figure}
\begin{center}
\hspace*{-0.04\textwidth}
\includegraphics[width=0.55\textwidth]{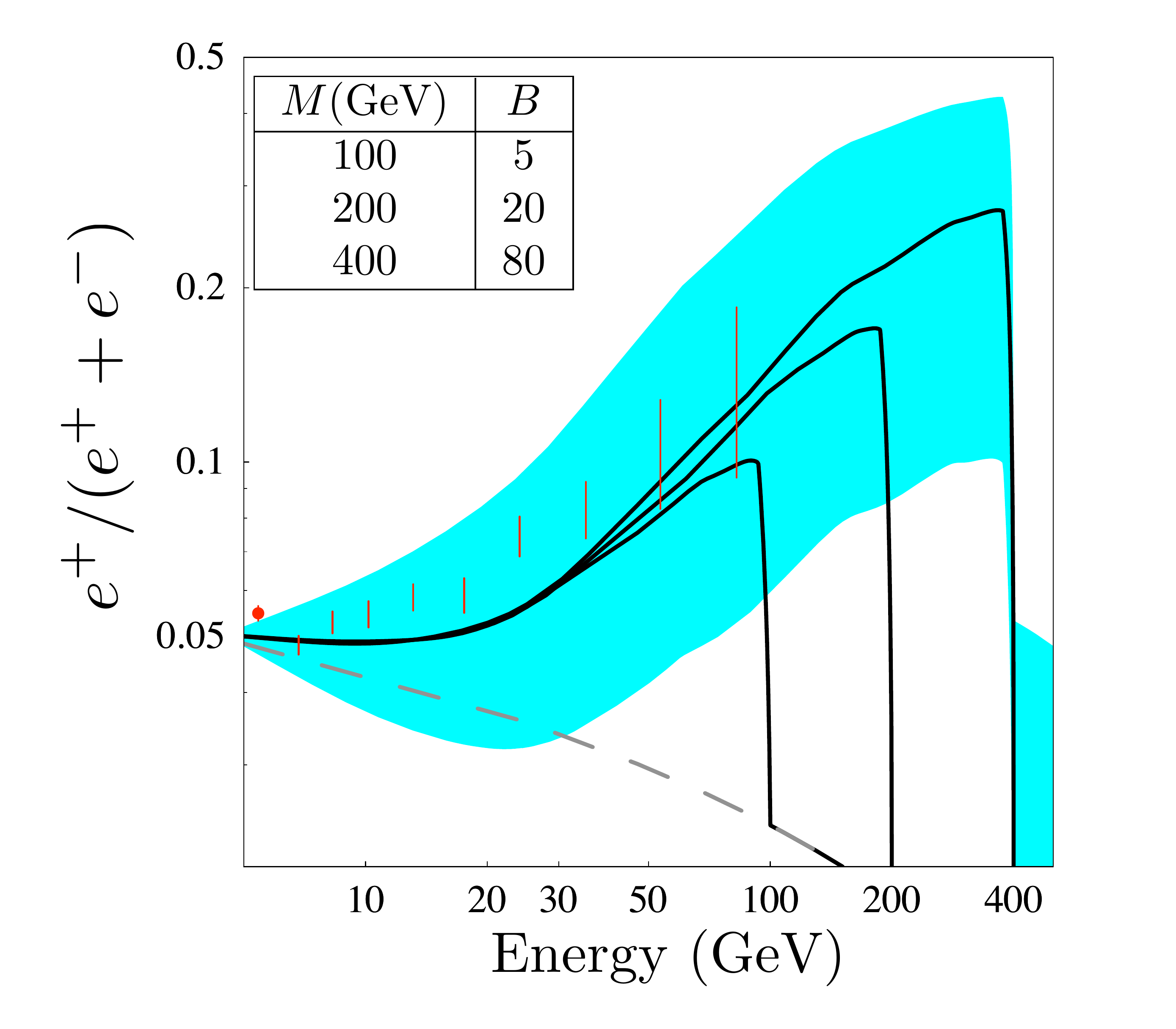}
\vspace*{-9mm}
\end{center}
\caption{Same Fig.~\ref{DC150-emutau}, for 
$M = 100,200,400$ GeV\@.  The DC model was used for
propagation, and annihilation was assumed only into $e^+e^-$.}
\label{eDC-masses}
\end{figure}

In Figs.~\ref{e150boost5} and \ref{e150boost15} we show
the predicted positron ratio spectra for $M = 150$ GeV
and boost factors of $5$ and $15$ respectively,
again showing the astrophysical uncertainty from
the propagation parameters as well as the electron flux.
In Fig.~\ref{DC150-emutau} we show a comparison between
annihilation to $e^+e^-$ final state only versus democratic 
to $\ell^+\ell^-$.  The boost factors were set to 
$10$ and $30$ respectively.
In Fig.~\ref{eDC-masses} we show 
the predicted positron ratio spectra for a range of
masses $M = 100,200,400$ GeV\@.  The corresponding boost
factors are $5$, $20$, and $80$.  Notice that the scaling
of the boost factor with mass is simply $B \propto M^2$.
The boost factor required for $M = 400$ GeV seems probably
beyond anything plausible from clumping.  Nevertheless,
if the ATIC/PPB-BETS hint persists, it is at least possible
to simultaneously explain both excesses.

One observational prediction is clear.  If a sharp drop in
the positron spectrum is observed by PAMELA, this would
provide strong evidence in favor of Dirac dark matter
with a small to modest boost factor.  It would suggest 
the ATIC/PPB-BETS excess is either an observational 
anomaly or unrelated to the PAMELA observations.
On the other hand, if PAMELA were to observe a continuous
rise in the positron fraction, this would provide evidence
that a more massive particle annihilating to $e^+e^-$ 
is simultaneously explaining both excesses.  This would
seem to require a larger boost factor.  Exactly how large
is dependent on the background flux of electrons.
For a spectrum falling as $E^{-3.5}$ ($E^{-3.15}$), 
the boost factor needs to be of order $16$ ($80$).
Our estimates suggest the best fit to the present data
in somewhat in between this range.  If future evidence 
is further strengthened for the ATIC/PPB-BETS excess, 
we will be able to make more precise statements about
the parameters and boost factors that best fit the data.

\section{The Dirac Bino as Dirac Dark Matter}

Our discussion up to now has been completely general with
respect to an effective theory of Dirac dark matter.
One candidate for Dirac dark matter seems particularly compelling:
a Dirac bino in a low energy supersymmetric model.
A Dirac bino arises in supersymmetry when the bino -- the fermionic 
superpartner contained in the hypercharge superfield strength $W_Y^\alpha$ -- 
acquires a Dirac mass with a gauge singlet $S$.
This occurs when supersymmetry breaking arises from $D$-terms
\cite{Fayet:1978qc,Polchinski:1982an,Hall:1990hq,Fox:2002bu}.
For the Dirac bino, the operator is
\begin{eqnarray}
\sqrt{2} \int d^2\theta \frac{W'_\alpha W_Y^\alpha S}{M_\star} \; 
\end{eqnarray}
which gives rise to a Dirac mass term 
\begin{eqnarray}
\frac{D'}{M_\star} (\lambda \psi + h.c.)
\end{eqnarray}
where $\lambda$ and $\psi$ are the 2-component bino and 
singlet fermions, respectively.
If no Majorana mass is generated by supersymmetry breaking,
this mass term implies the Dirac bino is a pure Dirac fermion.
This could be accomplished by accident (tuning all contributions
to the Majorana mass to conspire to vanish) or by symmetries
(supersymmetry breaking that respects a $U(1)_R$ symmetry,
for example).  In this section, we will first assume a 
Dirac bino exists, and consider the implications.
At the end we will consider a model in which a Dirac bino 
may be automatic.

\begin{figure}
\begin{center}
\includegraphics[width=0.45\textwidth]{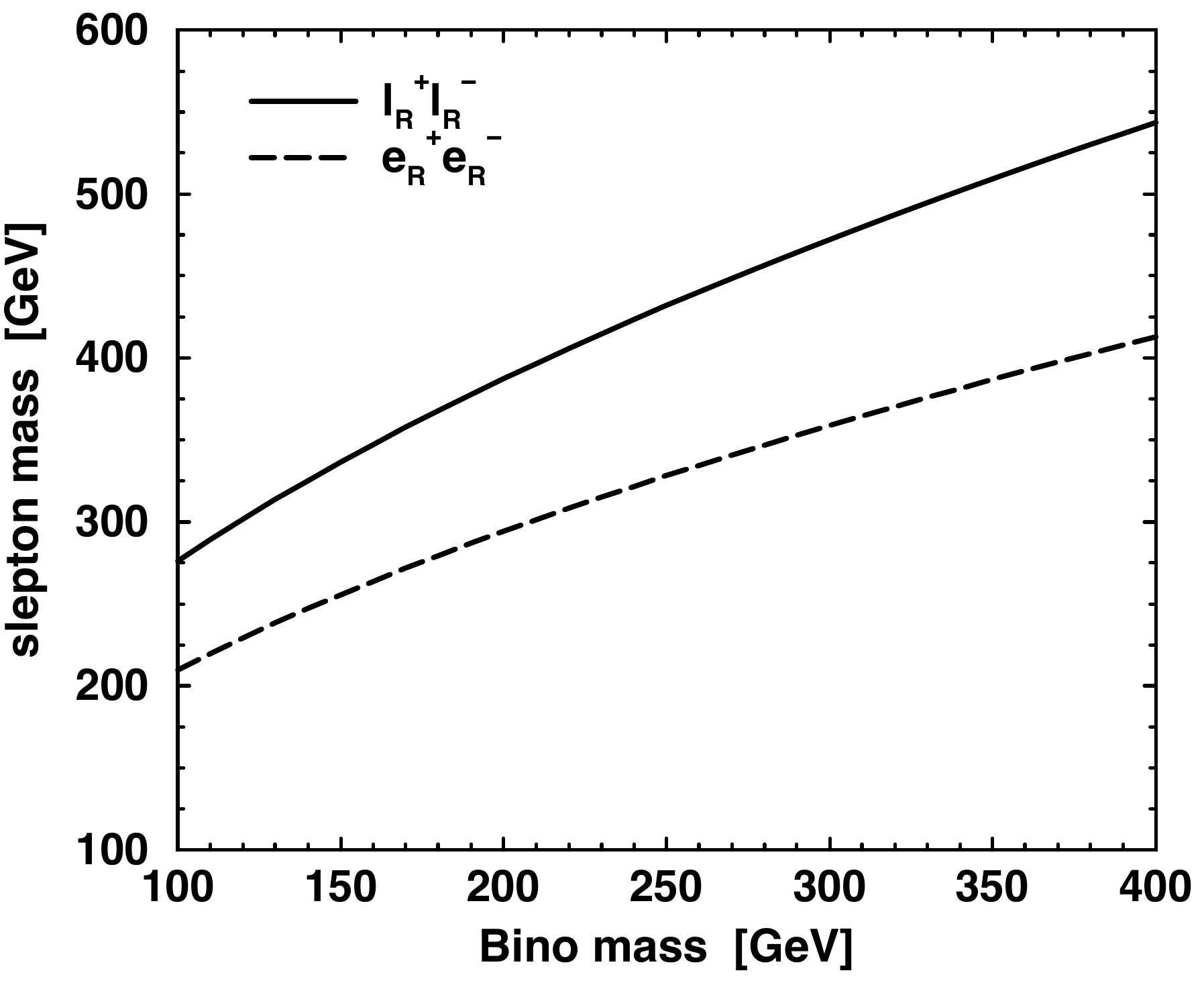}
\end{center}
\caption{Masses of the right-handed scalars such that
the Dirac bino has a thermal relic abundance, $\Omega h^2 = 0.114$,
consistent with cosmological data.
The top curve corresponds to the flavor-democratic
scenario, $m_{\tilde{e}_R} = m_{\tilde{\mu}_R} = m_{\tilde{\tau}_R}$,
while the lower curve corresponds to electrons only
$m_{\tilde{e}_R} = 1$.
In both cases we took only right-handed leptons 
for simplicity; adding left-handed leptons is trivial.}
\label{omega-fig}
\end{figure}

The relic abundance of an exact Dirac bino has been 
calculated before in Ref.~\cite{Hsieh:2007wq} using 
$t$-channel (and $u$-channel) scalar exchange.
Left-handed [right-handed] scalars give rise to a four-fermion 
interaction that can be Fierz transformed into our 
effective operators Eq.~(\ref{left-op})-(\ref{right-op}) 
with $c_L = (Y_L g')^2/2$ and $c_R = (Y_R g')^2/2$.
Here $Y_f$ is the hypercharge of the Standard Model fermions
and $g'$ is the hypercharge coupling.
The cutoff scale is the mass of the exchanged scalar 
$\Lambda = m_{\tilde{f}}$.  This allows us to immediately
re-evaluate Fig.~\ref{effomega-fig} in terms of the masses
of the physical scalar states that resolve the four-fermion 
operators.  This is shown in Fig.~\ref{omega-fig}.

The dominance of the leptonic operators becomes clear
for two reasons.  First, the four-fermion operators to any
Standard Model fermion are proportional to $Y_f^4$ (very much like 
KK dark matter \cite{Cheng:2002ej,Hooper:2004xn}),
which is largest for the right-handed leptons.
Second, since the operators scale as $1/m_f^4$, 
even a modest hierarchy in which sleptons are lighter than squarks 
will overwhelmingly cause the dominant annihilation channel to 
proceed through right-handed leptons.  Hence, a Dirac bino naturally 
explains annihilation to charged leptons.  
The collider implication is clear:  relatively light sleptons,
in a mass range between about $200$-$400$ GeV, are an
inescapable consequence to obtain a thermal relic abundance
consistent with cosmology and a positron signal consistent with
PAMELA\@.

A pure Dirac bino-eigenstate has no coupling to the $Z$.
This eliminates one source of vector interactions to quarks that 
would be devastating given the current nuclear recoil direct detection
bounds.  In the presence of Dirac Higgsinos, however, a residual
coupling to the $Z$ is generated.  We can estimate the size
of the $D$-$D$-$Z$ coupling using the mass-insertion approximation,
and we obtain $(g' v)^2/\mu^2$ 
where $v = 174$ GeV and $\mu$ is a Dirac Higgsino mass.  
This coupling needs to be smaller than about $0.01$ to be 
safe against direct detection bounds \cite{Belanger:2007dx}, 
and thus we obtain $\mu > 600$ GeV\@.  

A Dirac bino could also be seen in direct detection experiments
through the exchange of squarks.  The largest contribution
arises from the exchange of first-generation right-handed 
up-type squarks.  A preliminary estimate of the size of this 
contribution is that we need $m_{\tilde{u}_R} \gsim 1.5$ TeV\@.
The other first generation squarks can be proportionally 
lighter, the bounds scaling roughly with hypercharge per
exchange sfermion.

As we have seen from our model-independent analysis above,
the boost factor is minimized when the annihilation proceeds
only into electrons.  Again, a mild hierarchy between
$m_{\tilde{e}_R} < m_{\tilde{\mu}_R},m_{\tilde{\tau}_R}$
would suffice to explain this.  However, this is atypical 
of flavor-blind mediation mechanisms.  

An intriguing possibility is that all gauginos are Dirac fermions,
due to an exact $R$-symmetry that is preserved by
supersymmetry breaking.
This has been considered recently in \cite{Kribs:2007ac},
where it was shown that the squark and slepton masses could
be nearly flavor-arbitrary without violating the constraints
from flavor-violation in the Standard Model.
This is suggestive as an explanation of why the sleptons
could have a modest mass hierarchy while satisfying
bounds from lepton flavor violation.
However, the scenario we envision here is slightly different than 
was originally proposed in \cite{Kribs:2007ac}.
Here, the Dirac bino must be the lightest supersymmetric particle,
and the sleptons must be heavier than the bino by roughly a factor 
of $2$-$3$.  This is the opposite mass hierarchy from what was considered 
in \cite{Kribs:2007ac}, and while the parametric scaling of
the operators leading to lepton flavor violation is not expected
to change, the full extent of the allowed flavor mixings in the
slepton masses requires a re-evaluation of lepton flavor violation.
This is in progress and will be presented elsewhere.

In an $R$-symmetric model, the dimension-5 operator Eq.~(\ref{dim-5})
is absent since it is forbidden by the $R$-symmetry.
There are, however, new dimension-5 operators that are
allowed by the $R$-symmetry, such as 
$\overline{D} D \tilde{H}^\dagger \tilde{R}^\dagger$,
which are suppressed by the Dirac Higgsino mixing mass $\mu$.
Since $\mu$ gives rise to both a Dirac Higgsino mixing mass 
as well as an $\tilde{R}$ scalar mass, and it must be
larger than 600 GeV to keep the $D$-$D$-$Z$ coupling in line,
this operator is not kinematically relevant for the
Dirac masses we considered here.

Another important issue is to understand how well the $R$-symmetry 
may be preserved.  Supersymmetry breaking without $R$-symmetry breaking
is commonplace, but $R$-violation creeps back in,
due to the constant that is added to the superpotential of 
supergravity to cancel off the cosmological constant \cite{Bagger:1999rd}.
It is generally expected that this violation of $R$-symmetry
in the hidden sector would feed into the visible sector
through anomaly mediation.  This would lead to Majorana masses
for the gauginos, which would split the Dirac bino state into a 
two Majorana fermions.  The heavier state would then decay into the
lighter state on a timescale rapid compared to the age of the Universe,
unless the mass splitting is very small.
The resulting dark matter present in our galaxy today 
would be made of Majorana binos, which would not efficiently 
annihilate to leptons and thus would not be expected to explain
the PAMELA excess.
Dialing down the mass splitting is possible by lowering the
gravitino mass, but then this allows the bino to decay into a 
light gravitino.  
There may be possible escapes with supergravity \cite{Bagger:1999rd}
or otherwise postulating supersymmetry without supergravity
\cite{Luty:2002ff,Luty:2002hj}, 
though our UV theory becomes an effective theory that will break down 
at an intermediate scale.

\section{Discussion}

We have presented an effective theory of a stable Dirac fermion
dark matter candidate that couples to the Standard Model through 
dimension-6 four-fermion interactions.  The annihilation rate is
not velocity-suppressed, and thus the freeze-out thermally-averaged
annihilation cross section is the same as the annihilation 
cross section that occurs in our galactic neighborhood.
We have shown that if the annihilation proceeds into
right-handed electrons, the PAMELA positron ratio excess can 
be explained with a minimal boost factor.  Given present
astrophysical uncertainties, the boost factor could be as
small as $1$ for $M = 100$ GeV if either 
$\rho_{8.5} = 0.7$~GeV/cm$^3$ and $\Phi_{e^-} \propto E^{-3.15}$
or $\rho_{8.5} = 0.3$~GeV/cm$^3$ and $\Phi_{e^-} \propto E^{-3.5}$.
Larger Dirac fermion masses are possible to the extent that 
larger boost factors are plausible, scaling as $B \propto M^2$.  

If the mass of the Dirac fermion is within the energy range
that PAMELA can explore, they should see a 
striking feature in the positron fraction at the mass of
the Dirac fermion.  Fermi/GLAST could also see a feature
in their photon spectrum resulting from final state radiation
off one of the charged leptons.  If the mass is large,
this could simultaneously explain the hint of an excess at
ATIC and PPB-BETS with a boost factor as small as $16$ given
the present uncertainty in the electron spectra.

Finally, we showed that a natural candidate for a Dirac fermion
dark matter particle is a Dirac bino in supersymmetry.
This can automatically explain the dominance of the 
right-handed leptonic operators.  The reminder of the
spectrum is somewhat restricted by the constraints from
nuclear recoil direct detection.  A detailed analysis of the
model and spectrum is in progress.
The inevitable consequence of this interpretation is 
the presence of relatively light sleptons, with masses $200$-$400$ GeV,
the range depending on the precise mass of the Dirac bino
and the number of flavors of sleptons that are light.  
Minimizing the boost factor implies the right-handed selectron 
is the lightest slepton, which is characteristic
of our model.  
This provides an exciting opportunities for LHC\@.

As a final concluding thought, is it possible to reconcile
the PAMELA excess with the observation by DAMA/LIBRA 
of an annual modulation in nuclear recoil \cite{Bernabei:2008yi}?
In an effective theory, it is straightforward to 
implement the inelastic mechanism 
\cite{TuckerSmith:2001hy,TuckerSmith:2002af,Chang:2008gd}
to explain the annual modulation.
A second Dirac fermion $D'$ is added to the effective theory, 
with $D$ and $D'$ permitted to mix with each other
under a global $U(1)_D$ conservation.   The heavier mass eigenstate
is taken to be about 100 keV heavier than the actual Dirac dark matter
particle.  Then, one can simply add a four-fermion dimension-6
quark operator to the effective theory,
$\dbar' \gamma^\mu D \overline{q} \gamma_\mu q/\Lambda^2$.
The key is to ensure that a $\dbar' D$ or
$\dbar D'$ operator is generated while a $\dbar D$
operator to quarks is not.
This means a vector interaction exists between the light state 
$D$ to the excited state $D'$ with quarks of a scattered nucleus.
Realizing this pattern of operators in a model is an interesting 
(and challenging) model-building problem we leave for future study.
Interestingly, the range of dark matter masses that most easily 
permit an inelastic explanation of the annual modulation is
about $70-250$ GeV\@.  This exactly coincides with the range
that is most favorable towards an explanation of the PAMELA
positron ratio excess.

\begin{acknowledgments} 

The authors thank I.~Moskalenko and A.~Strong for help in
understanding the physics and output of their Galprop program;
J.~Schombert for teaching us how to read FITS files;
and N.~Weiner for useful discussions at an early stage
in the project.
The authors also thank the Aspen Center for Physics where this 
work was initiated.  
This work was supported in part by the Department of Energy under 
grant numbers DE-AC02-76SF00515 (RH) and DE-FG02-96ER40969 (GDK).

\end{acknowledgments}


\begin{thebibliography}{99}


\bibitem{Adriani:2008zr}
  O.~Adriani {\it et al.},
  arXiv:0810.4995 [astro-ph].

\bibitem{Cheng:2002ej}
  H.~C.~Cheng, J.~L.~Feng and K.~T.~Matchev,
  Phys.\ Rev.\ Lett.\  {\bf 89}, 211301 (2002)
  [arXiv:hep-ph/0207125].

\bibitem{Hooper:2004xn}
  D.~Hooper and G.~D.~Kribs,
  Phys.\ Rev.\  D {\bf 70}, 115004 (2004)
  [arXiv:hep-ph/0406026].

\bibitem{Hooper:2004bq}
  D.~Hooper and J.~Silk,
  Phys.\ Rev.\  D {\bf 71}, 083503 (2005)
  [arXiv:hep-ph/0409104].

\bibitem{Cirelli:2008id}
  M.~Cirelli, R.~Franceschini and A.~Strumia,
  Nucl.\ Phys.\  B {\bf 800}, 204 (2008)
  [arXiv:0802.3378 [hep-ph]].

\bibitem{Bergstrom:2008gr}
  L.~Bergstrom, T.~Bringmann and J.~Edsjo,
  arXiv:0808.3725 [astro-ph].

\bibitem{Cirelli:2008jk}
  M.~Cirelli and A.~Strumia,
  arXiv:0808.3867 [astro-ph].

\bibitem{Barger:2008su}
  V.~Barger, W.~Y.~Keung, D.~Marfatia and G.~Shaughnessy,
  arXiv:0809.0162 [hep-ph].

\bibitem{Cholis:2008hb}
  I.~Cholis, L.~Goodenough, D.~Hooper, M.~Simet and N.~Weiner,
  arXiv:0809.1683 [hep-ph].

\bibitem{Cirelli:2008pk}
  M.~Cirelli, M.~Kadastik, M.~Raidal and A.~Strumia,
  arXiv:0809.2409 [hep-ph].

\bibitem{Huh:2008vj}
  J.~H.~Huh, J.~E.~Kim and B.~Kyae,
  arXiv:0809.2601 [hep-ph].

\bibitem{ArkaniHamed:2008qn}
  N.~Arkani-Hamed, D.~P.~Finkbeiner, T.~Slatyer and N.~Weiner,
  arXiv:0810.0713 [hep-ph].

\bibitem{ArkaniHamed:2008qp}
  N.~Arkani-Hamed and N.~Weiner,
  arXiv:0810.0714 [hep-ph].

\bibitem{Finkbeiner:2008qu}
  D.~P.~Finkbeiner, T.~Slatyer and N.~Weiner,
  arXiv:0810.0722 [hep-ph].

\bibitem{Pospelov:2008jd}
  M.~Pospelov and A.~Ritz,
  arXiv:0810.1502 [hep-ph].

\bibitem{Hooper:2008kg}
  D.~Hooper, P.~Blasi and P.~D.~Serpico,
  arXiv:0810.1527 [astro-ph].

\bibitem{Hisano:2008ti}
  J.~Hisano, M.~Kawasaki, K.~Kohri and K.~Nakayama,
  arXiv:0810.1892 [hep-ph].

\bibitem{Yuksel:2008rf}
  H.~Yuksel, M.~D.~Kistler and T.~Stanev,
  arXiv:0810.2784 [astro-ph].

\bibitem{Kamionkowski:2008gj}
  M.~Kamionkowski and S.~Profumo,
  arXiv:0810.3233 [astro-ph].

\bibitem{Khalil:2008ps}
  S.~Khalil and H.~Okada,
  arXiv:0810.4573 [hep-ph].

\bibitem{Serpico:2008te}
  P.~D.~Serpico,
  arXiv:0810.4846 [hep-ph].

\bibitem{Nelson:2008hj}
  A.~E.~Nelson and C.~Spitzer,
  arXiv:0810.5167 [hep-ph].

\bibitem{Donato:2008jk}
  F.~Donato, D.~Maurin, P.~Brun, T.~Delahaye and P.~Salati,
  arXiv:0810.5292 [astro-ph].

\bibitem{Cholis:2008xx}
  I.~Cholis, D.~P.~Finkbeiner, L.~Goodenough and N.~Weiner,
  arXiv:0810.5344 [astro-ph].

\bibitem{Nomura:2008xx}
  Y.~Nomura and J.~Thaler
  arXiv:0810.5397v1 [hep-ph]

\bibitem{Komatsu:2008hk}
  E.~Komatsu {\it et al.}  [WMAP Collaboration],
  arXiv:0803.0547 [astro-ph].

\bibitem{Moskalenko:1997gh}
  I.~V.~Moskalenko and A.~W.~Strong,
  Astrophys.\ J.\  {\bf 493}, 694 (1998)
  [arXiv:astro-ph/9710124].

\bibitem{galprop}
  http://galprop.stanford.edu/web\_galprop/galprop\_home.html;
  A.~W.~Strong and I.~V.~Moskalenko,
  arXiv:astro-ph/9906228;
  A.~W.~Strong and I.~V.~Moskalenko,
  arXiv:astro-ph/0106504.

\bibitem{Alcaraz:2000bf}
  J.~Alcaraz {\it et al.}  [AMS Collaboration],
  Phys.\ Lett.\  B {\bf 484}, 10 (2000)
  [Erratum-ibid.\  B {\bf 495}, 440 (2000)].

\bibitem{ATIC}
  ATIC Collaboration, ICRC 2005, 
  http://icrc2005.tifr.res.in/htm/Vol-Web/Vol-13/13001-chn-chang-J-abs1-og11-oral.pdf.

\bibitem{Torii:2001aw}
  S.~Torii {\it et al.},
  Astrophys.\ J.\  {\bf 559}, 973 (2001).

\bibitem{Torii:2008xu}
  S.~Torii {\it et al.},
  arXiv:0809.0760 [astro-ph].

\bibitem{CAPRICE}
  M.~Boezio {\it et al.}, 
  Astrophys.\ J.\  {\bf 532}, 653 (2000)

\bibitem{DuVernois:2001bb}
  M.~A.~DuVernois {\it et al.},
  Astrophys.\ J.\  {\bf 559}, 296 (2001).

\bibitem{Grimani:2002yz}
  C.~Grimani {\it et al.},
  Astron.\ Astrophys.\  {\bf 392}, 287 (2002).

\bibitem{pamelaSLAC}
  E.~Vannuccini (PAMELA Collaboration), 
  ``Recent Results from PAMELA'', SLAC Summer Institute (2008);
  http://www-conf.slac.stanford.edu/ssi/2008/080708t\_Vannuccini.pdf.

\bibitem{Birkedal:2005ep}
  A.~Birkedal, K.~T.~Matchev, M.~Perelstein and A.~Spray,
  arXiv:hep-ph/0507194.

\bibitem{Kobayashi:1999he}
  T.~Kobayashi {\it et al.},
{\it Prepared for 26th International Cosmic Ray Conference (ICRC 99), Salt Lake City, Utah, 17-25 Aug 1999}

\bibitem{Belanger:2007dx}
  G.~Belanger, A.~Pukhov and G.~Servant,
  JCAP {\bf 0801}, 009 (2008)
  [arXiv:0706.0526 [hep-ph]].

\bibitem{Kribs:2007ac}
  G.~D.~Kribs, E.~Poppitz and N.~Weiner,
  arXiv:0712.2039 [hep-ph].

\bibitem{Ahmed:2008eu}
  Z.~Ahmed {\it et al.}  [CDMS Collaboration],
  arXiv:0802.3530 [astro-ph].

\bibitem{Angle:2007uj}
  J.~Angle {\it et al.}  [XENON Collaboration],
  Phys.\ Rev.\ Lett.\  {\bf 100}, 021303 (2008)
  [arXiv:0706.0039 [astro-ph]].

\bibitem{Adriani:2008zq}
  O.~Adriani {\it et al.},
  arXiv:0810.4994 [astro-ph].

\bibitem{Bernabei:2007gr}
  R.~Bernabei {\it et al.},
  Phys.\ Rev.\  D {\bf 77}, 023506 (2008)
  [arXiv:0712.0562 [astro-ph]].

\bibitem{Srednicki:1988ce}
  M.~Srednicki, R.~Watkins and K.~A.~Olive,
  Nucl.\ Phys.\  B {\bf 310}, 693 (1988).

\bibitem{Hsieh:2007wq}
  K.~Hsieh,
  Phys.\ Rev.\  D {\bf 77}, 015004 (2008)
  [arXiv:0708.3970 [hep-ph]].

\bibitem{Goldberg:1983nd}
  H.~Goldberg,
  Phys.\ Rev.\ Lett.\  {\bf 50}, 1419 (1983).

\bibitem{Ambrosanio:1995it}
  S.~Ambrosanio, B.~Mele, G.~Montagna, O.~Nicrosini and F.~Piccinini,
  Nucl.\ Phys.\  B {\bf 478}, 46 (1996)
  [arXiv:hep-ph/9601292].

\bibitem{Baltz:1998xv}
  E.~A.~Baltz and J.~Edsjo,
  Phys.\ Rev.\  D {\bf 59}, 023511 (1999)
  [arXiv:astro-ph/9808243].

\bibitem{Casolino:2008zm}
  M.~Casolino and P.~Collaboration,
  arXiv:0810.4980 [astro-ph].

\bibitem{Casadei:2004sb}
  D.~Casadei and V.~Bindi,
  Astrophys.\ J.\  {\bf 612}, 262 (2004).

\bibitem{Lionetto:2005jd}
  A.~M.~Lionetto, A.~Morselli and V.~Zdravkovic,
  JCAP {\bf 0509}, 010 (2005)
  [arXiv:astro-ph/0502406].

\bibitem{Amsler:2008zz}
  C.~Amsler {\it et al.}  [Particle Data Group],
  Phys.\ Lett.\  B {\bf 667}, 1 (2008).

\bibitem{Delahaye:2008ua}
  T.~Delahaye {\it et al.},
  arXiv:0809.5268 [astro-ph].

\bibitem{Bergstrom:1997fj}
  L.~Bergstrom, P.~Ullio and J.~H.~Buckley,
  Astropart.\ Phys.\  {\bf 9}, 137 (1998)
  [arXiv:astro-ph/9712318].

\bibitem{Diemand:2005vz}
  J.~Diemand, B.~Moore and J.~Stadel,
  Nature {\bf 433}, 389 (2005)
  [arXiv:astro-ph/0501589].

\bibitem{Lavalle:1900wn}
  J.~Lavalle, Q.~Yuan, D.~Maurin and X.~J.~Bi,
  arXiv:0709.3634 [astro-ph];
  J.~Lavalle, E.~Nezri, F.~S.~Ling, L.~Athanassoula and R.~Teyssier,
  arXiv:0808.0332 [astro-ph].

\bibitem{Gondolo:2005we}
  P.~Gondolo, J.~Edsjo, P.~Ullio, L.~Bergstrom, M.~Schelke and E.~A.~Baltz,
  New Astron.\ Rev.\  {\bf 49}, 149 (2005).

\bibitem{Fayet:1978qc}
  P.~Fayet,
  Phys.\ Lett.\  B {\bf 78}, 417 (1978).

\bibitem{Polchinski:1982an}
  J.~Polchinski and L.~Susskind,
  Phys.\ Rev.\  D {\bf 26}, 3661 (1982).

\bibitem{Hall:1990hq}
  L.~J.~Hall and L.~Randall,
  Nucl.\ Phys.\  B {\bf 352}, 289 (1991).

\bibitem{Fox:2002bu}
  P.~J.~Fox, A.~E.~Nelson and N.~Weiner,
  JHEP {\bf 0208}, 035 (2002)
  [arXiv:hep-ph/0206096].

\bibitem{Bagger:1999rd}
  J.~A.~Bagger, T.~Moroi and E.~Poppitz,
  JHEP {\bf 0004}, 009 (2000)
  [arXiv:hep-th/9911029].

\bibitem{Luty:2002ff}
  M.~A.~Luty,
  Phys.\ Rev.\ Lett.\  {\bf 89}, 141801 (2002)
  [arXiv:hep-th/0205077].

\bibitem{Luty:2002hj}
  M.~A.~Luty and N.~Okada,
  JHEP {\bf 0304}, 050 (2003)
  [arXiv:hep-th/0209178].

\bibitem{Bernabei:2008yi}
  R.~Bernabei {\it et al.}  [DAMA Collaboration],
  Eur.\ Phys.\ J.\  C {\bf 56}, 333 (2008)
  [arXiv:0804.2741 [astro-ph]].

\bibitem{TuckerSmith:2001hy}
  D.~Tucker-Smith and N.~Weiner,
  Phys.\ Rev.\  D {\bf 64}, 043502 (2001)
  [arXiv:hep-ph/0101138].

\bibitem{TuckerSmith:2002af}
  D.~Tucker-Smith and N.~Weiner,
  Nucl.\ Phys.\ Proc.\ Suppl.\  {\bf 124}, 197 (2003)
  [arXiv:astro-ph/0208403].

\bibitem{Chang:2008gd}
  S.~Chang, G.~D.~Kribs, D.~Tucker-Smith and N.~Weiner,
  arXiv:0807.2250 [hep-ph].




\end{thebibliography}
\end{document}